\documentclass[12pt,preprint]{aastex}

\begin{document}

\title{The Surprising Absence of Absorption in the Far-Ultraviolet
  Spectrum of Mrk~231}

\author{S. Veilleux\altaffilmark{1,2,3,4}, M.  Trippe\altaffilmark{1},
  F. Hamann\altaffilmark{5}, D. S. N. Rupke\altaffilmark{6},
  T. M. Tripp\altaffilmark{7}, H. Netzer\altaffilmark{8},
  D. Lutz\altaffilmark{4}, K. R. Sembach\altaffilmark{9},
  H. Krug\altaffilmark{1}, S. H. Teng\altaffilmark{10},
  R. Genzel\altaffilmark{4}, R. Maiolino\altaffilmark{11,12},
  E. Sturm\altaffilmark{4}, and L. Tacconi\altaffilmark{4}}

\altaffiltext{1}{Department of Astronomy, University of Maryland,
  College Park, MD 20742, USA; hkrug@astro.umd.edu,
  veilleux@astro.umd.edu, trippe@astro.umd.edu}

\altaffiltext{2}{Joint Space-Science Institute, University of Maryland,
  College Park, MD 20742, USA}

\altaffiltext{3}{Astroparticle Physics Laboratory, NASA Goddard Space
  Flight Center, Greenbelt, MD 20771, USA}

\altaffiltext{4}{Max-Planck-Institut f\"ur extraterrestrische
  Physik, Postfach 1312, D-85741 Garching, Germany}

\altaffiltext{5}{Department of Astronomy, University of Florida,
  Gainesville, FL 32611, USA}

\altaffiltext{6}{Department of Physics, Rhodes College, Memphis, TN 38112, USA}

\altaffiltext{7}{Department of Astronomy, University of Massachussetts,
  Amherst, MA 01003, USA}

\altaffiltext{8}{School of Physics and Astronomy and the Wise
  Observatory, The Raymond and Beverly Sackler Faculty of Exact
  Sciences, Tel-Aviv University, Tel-Aviv 69978 Israel}

\altaffiltext{9}{Space Telescope Science Institute, Baltimore, MD 21218, USA}

\altaffiltext{10}{Observational Cosmology Laboratory, NASA Goddard Space
  Flight Center, Greenbelt, MD 20771, USA}

\altaffiltext{11}{Cavendish Laboratory, University of Cambridge, 19
  J.J. Thomson Ave., Cambridge, CB3 0HE, UK}

\altaffiltext{12}{Kavli Institute for Cosmology, Madingley Road,
  Cambridge, CB3 0HA, UK}

\begin{abstract}
  Mrk~231, the nearest ($z$ = 0.0422) quasar, hosts both a
  galactic-scale wind and a nuclear-scale iron low-ionization broad
  absorption line (FeLoBAL) outflow. We recently obtained a
  far-ultraviolet (FUV) spectrum of this object covering $\sim$1150 --
  1470 \AA\ with the Cosmic Origins Spectrograph on board the {\em
    Hubble Space Telescope}. This spectrum is highly peculiar,
  highlighted by the presence of faint ($\la$2\% of predictions based
  on H$\alpha$), broad ($\ga$ 10,000 km s$^{-1}$ at the base), and
  highly blueshifted (centroid at $\sim$ --3500 km s$^{-1}$)
  Ly$\alpha$ emission.  The FUV continuum emission is slightly
  declining at shorter wavelengths (consistent with $F_\lambda \propto
  \lambda^{1.7}$) and does not show the presence of any obvious
  photospheric or wind stellar features. Surprisingly, the FUV
  spectrum also does not show any unambiguous broad absorption
  features. It thus appears to be dominated by the AGN, rather than
  hot stars, and virtually unfiltered by the dusty FeLoBAL screen.
  The observed Ly$\alpha$ emission is best explained if it is produced
  in the outflowing BAL cloud system, while the Balmer lines arise
  primarily from the standard broad emission line region seen through
  the dusty ($A_V \sim 7$ mag.) broad absorption line region.  Two
  possible geometric models are discussed in the context of these new
  results.
\end{abstract}


\section{Introduction}

At a distance$\footnote{Based on a redshift $z$ = 0.0422 and a
  cosmology with $H_0$ = 73 km s$^{-1}$ Mpc$^{-1}$, $\Omega_{\rm
    matter}$ = 0.27, and $\Omega_{\rm vacuum}$ = 0.73.}$ of only 178
Mpc, where 1$\arcsec$ = 0.863 kpc, Mrk~231 is the nearest quasar known
(Boksenberg et al. 1977). The analysis by Braito et al. (2004) of the
joint {\em XMM-Newton} -- {\em BeppoSAX} X-ray spectrum of this object
has revealed the existence of a highly absorbed ($N_H$ $\sim$ 2
$\times$ 10$^{24}$ cm$^{-2}$), powerful (intrinsic 2-10 keV luminosity
of $\sim$ 1 $\times$ 10$^{44}$ erg s$^{-1}$) AGN.  Mrk~231 is also an
ultraluminous infrared galaxy (ULIRG) with infrared (8 -- 1000 $\mu$m)
luminosity log[$L_{\rm IR}$/$L_\odot$] = 12.54 and a bolometric
luminosity log[$L_{\rm BOL}$/$L_\odot$] $\approx$
12.60.$\footnote{This estimate for the bolometric luminosity includes
  all IR and non-IR contributions and is derived by simply assuming
  $L_{\rm BOL}$ = 1.15 L$_{\rm IR}$, typical for ULIRGs (e.g., Sanders
  \& Mirabel 1996).}$ A detailed analysis of its {\em Spitzer}
mid-infrared spectrum indicates that $\sim$30\% of this bolometric
luminosity is produced by a circumnuclear starburst with SFR $\sim$
170 M$_\odot$ yr$^{-1}$, surrounding the dominant optical broad-lined
quasar (Veilleux et al. 2009). The disturbed morphology of the host
suggests that both the quasar and starburst are triggered by a recent
merger event (e.g., Hamilton \& Keel 1987; Hutchings \& Neff 1987;
Surace et al. 1998; Veilleux et al. 2002, 2006).

In recent years, Mrk~231 has become the archetype of galactic-scale
quasar-driven winds. These outflow events are purported to
self-regulate the growth of the black hole (BH) and spheroidal
component of the galaxy, and explain the relatively tight BH-spheroid
mass relation (e.g., Di Matteo et al. 2005; Murray et al. 2005;
Veilleux, Cecil, \& Bland-Hawthorn 2005 and references
therein). Evidence in Mrk~231 for a powerful kpc-scale wind with
outflow velocity of up to $\sim$1000 km s$^{-1}$ and mass outflow rate
of order $\sim$ 400 -- 1000 M$_\odot$ yr$^{-1}$ has now been seen in
the neutral gas (Rupke, Veilleux, \& Sanders 2005; Rupke \& Veilleux
2011, 2012), the molecular gas (Feruglio et al. 2010; Fischer et
al. 2010; Sturm et al. 2011; Aalto et al. 2012; Cicone et al. 2012),
and the warm ionized gas (Lipari et al. 2009; Rupke \& Veilleux 2011,
2012).  It is not clear whether this galactic-scale wind is physically
related to the well-known unresolved broad absorption-line (BAL)
systems in the core of this object (Boksenberg et al. 1977; Rudy,
Foltz, \& Stocke 1985; Hutchings \& Neff 1987; Boroson et al. 1991;
Kollatschny, Dietrich, \& Hagen 1992; Forster, Rich, \& McCarthy 1995;
Smith et al. 1995; Rupke, Veilleux, \& Sanders 2002; Gallagher et
al. 2002, 2005). The Mrk~231 BAL features are detected in several
transitions including many low-ionization species (e.g., Na I D
$\lambda\lambda$5890, 5896, He I$^*$ $\lambda$3889, Ca II H and K, and
Mg II $\lambda$2800, as well as Fe II UV1 $\lambda$2599 and Fe II UV2
$\lambda$2382, hence fitting the rare FeLoBAL category). They show
strong absorption across the range $\sim$ --3500 to --5500 km s$^{-1}$
and no significant absorption at velocities above --3500 km s$^{-1}$
(see, e.g., Rupke et al. 2002).

Mrk~231 was the first target observed under our far-ultraviolet (FUV)
spectroscopic survey of the nearby {\em QUEST} ({\em Q}uasar -- {\em
  U}LIRG {\em E}volution {\em St}udy; e.g., Veilleux 2012 and
references therein) QSOs with {\em HST} (PID \#12569, PI
Veilleux). The unprecedented sensitivity of the Cosmic Origins
Spectrograph (COS, Green et al. 2012) allowed us to obtain the first
high signal-to-noise (S/N) ratio spectrum of Mrk~231 below $\sim$1500
\AA. This paper describes the results from our analysis of this
spectrum.  The results from the survey will be presented in future
papers. The acquisition and reduction of the COS data and
complementary optical spectrum of Mrk~231 are discussed in Section 2,
followed by a description of the results (Section 3) and a discussion
of the implications (Section 4). Throughout this paper we use $z$ =
0.0422 as the rest-frame velocity of Mrk~231; this value is considered
quite secure since it is based on rotation curves derived from the
ionized, neutral, and molecular gas components and the peak of the H~I
21-cm absorption feature (see Rupke \& Veilleux 2011 for references).

\section{Observations}

\subsection{{\em HST}--COS FUV Spectroscopy}

Mrk~231 was observed for a total 12,539 seconds spread over 5 orbits
using the G130M grating and the 2$\farcs$5 $\times$ 2$\farcs$5
aperture to cover $\sim$1150 -- 1470 \AA\ at a spectral resolution of
$\sim$15-20 km s$^{-1}$.  All of these observations were done in
TIME-TAG mode to provide temporal sampling and allow us to exclude
poor quality data and get improved thermal correction and background
removal. In the end, we did not have to exclude any data for
Mrk~231. To reduce the fixed pattern noise and fill up the wavelength
hole produced by the chip gap without excessive overheads, we split
each orbit into four separate segments of similar durations at two
different FP$\_$POS offset positions (either \#1 and \#3 or \#2 and
\#4, depending on the orbit) and two different central wavelengths
(1309 or 1327 \AA).  All observations were carried out on 15 October
2011 and processed using the standard {\em calcos}$\footnote{See
  Chapter 3 of Part II of the HST Data Handbook for COS
  (http://www.stsci.edu/hst/cos/documents/handbooks/datahandbook) for
  the details of the calcos processing.}$ calibration
pipeline. Flatfielding, alignment, and co-addition of the individual
exposures were performed using the IDL routines provided by the COS
GTO team.$\footnote{Routines are available at
  http://casa.colorado.edu/$\sim$danforth/science/cos/costools.html,
  and described in Danforth et al. (2010) and Keeney et al. (2012).}$

\subsection{Ground-based Optical Spectroscopy}

The highest-velocity ($\sim$ --8000 km s$^{-1}$) component of the BAL
systems in Mrk~231 has proven to be time-variable (Hutchings \& Neff
1987; Boroson et al.\ 1991; Kollatschny et al. 1992; Forster et
al. 1995; Rupke et al. 2002, 2005), so we decided to get a new optical
spectrum of this object to complement our 2001, 2004, and 2007 spectra
published in Rupke et al. (2002), Rupke et al. (2005c), and Rupke \&
Veilleux (2011), respectively, and facilitate the task of comparing
these data with the COS FUV spectrum. The instrumental setup used for
these ground-based observations was the same as that used for the 2004
optical spectra: the Ritchey-Chr\'etien spectrograph on the Mayall
4-meter telescope at Kitt Peak was used on 27 April 2012 with grating
KPC-18C in the first order and a 1$\farcs$25 slit to obtain 4 spectra
each of 15 min duration with a spectral resolution of 85 km
s$^{-1}$. Standard IRAF$\footnote{http://iraf.noao.edu/}$ procedures
were used to reduce and calibrate this spectrum.

\section{Results}

A near-ultraviolet (NUV) COS image of Mrk~231, obtained for accurate
target acquisition, is consistent with an unresolved point source
centered on the optical position of the quasar at the resolution of
the COS imager (FWHM $\sim$ 0$\farcs$05).  The FUV spectrum of Mrk~231
is shown in Figures 1 and 2 and is highly peculiar.  We have been able
to identify unambiguously only one emission feature intrinsic to
Mrk~231: broad, highly blueshifted Ly$\alpha$ emission (discussed in
detail in the last paragraph of this Section). This lack of FUV
features contrasts starkly with the rich optical spectrum of this
object (shown in Figure 3) and with the rest-frame UV spectra of other
FeLoBAL quasars (e.g., Hall et al. 2002). These and other quasar
spectra indicate that Mrk~231 should have strong broad absorption in
Ly$\alpha$ and possibly also emission and absorption from P~V
$\lambda\lambda$1118, 1128, Si~III $\lambda$1206, N~V
$\lambda\lambda$1238, 1243, Si~II $\lambda$1260 (Si~II$^*$
$\lambda$1265), OI $\lambda$1302, Si~II $\lambda$1304 (Si~II$^*$
$\lambda$1309), C~II $\lambda$1335 (C~II$^*$ $\lambda$1336), and Si~IV
$\lambda\lambda$1394, 1403 within our spectral coverage. However, none
of these features is clearly detected. The peculiarity of the UV
spectrum of Mrk~231 was first noted by Hutchings \& Neff (1987) and
Lipari, Colina, \& Macchetto (1994) based on {\em IUE} spectra. It was
further investigated by Smith et al. (1995) using NUV FOS
spectropolarimetric data and by Gallagher et al. (2002) using a short
FOS calibration observation in the archives. In particular, Gallagher
et al. (2002) noted the likely presence of broad C~IV
$\lambda\lambda$1548, 1550 emission affected by a weak BAL, but the
S/N ratio of the FOS data was insufficient to provide firm limits on
the equivalent widths of the emission and absorption features below
$\sim$1450 \AA. The new COS data show that the spectrum of Mrk~231 at
$<$1450 \AA\ is also highly unusual.  As discussed in \S 4.3, we
believe that this is the result of a complex geometry with a dusty
broad absorption line region (BALR) obscuring portions of both the
continuum source and standard broad emission line region (BELR).

While the COS spectrum of Mrk~231 is dominated by Ly$\alpha$ emission,
it does also present a number of faint but significant ``bumps'' and
``wiggles''. We are confident that these features are real since they
are seen in individual exposures with different FP$\_$POS and central
wavelengths during each of the five orbits, but they are not seen in
the FUV spectra of our other targets, which were observed in the same
fashion reaching similar S/N ratios. Table 1 lists the properties of
these weak features along with tentative identifications (IDs).  None
of these IDs is secure except for the narrow absorption feature at
1252 \AA\, which is most likely a weak narrow H~I absorber at $\sim$
$-$3530 km s$^{-1}$ associated with either the nuclear outflow in
Mrk~231 or IGM along our line of sight. The significant width
($\sim$2000 km s$^{-1}$) of the absorption feature at $\sim$1392 \AA\
suggests an AGN origin.  If this feature forms in the $\sim$--4500 km
s$^{-1}$ BAL outflow seen in the optical Na~I, Ca~II, and He~I lines,
then the rest wavelength of the 1392 \AA\ feature should be $\sim$1356
\AA, for which there is no plausible ID.  The strongest absorption
line we expect to find near these wavelengths is C~II $\lambda$1335 (+
C~II$^*$ $\lambda$1336), but this implies much lower velocities, only
+200 (--70) km s$^{-1}$, compared to the optical BALs. Nonetheless,
this ID for the 1392 \AA\ feature is consistent with the weaker but
similar-looking absorption dip at $\sim$1452 \AA, which might be Si~IV
$\lambda$1394 at a similar near-systemic velocity. The apparent
absence of absorption in the other half of the Si~IV doublet, at 1403
\AA, is consistent with its factor of 2 lower f-value and the lower
S/N ratio in this part of the measured spectrum. The power-law fit in
Figure 1 also suggests the presence of shallow absorption features
around $\sim$1355 \AA\ ($\sim$1300 \AA\ in the rest-frame) and
$\sim$1225 \AA\ ($\sim$1176 \AA; partly affected by the damping wings
of Galactic Ly$\alpha$ absorption).  This latter feature may be
stellar photospheric C~III $\lambda$1176 near systemic velocity; we
return to this identification in \S 4.1.

The three emission features at $\sim$1181, 1296, and 1375 \AA\ could
not be matched unambiguously to redshifted atomic transitions expected
from Mrk~231 (or the ``template'' narrow-line Seyfert 1; e.g., Laor et
al. 1997).  The feature at 1181 \AA\ may be N~I $\lambda$1134
blueshifted by $\sim$ $-$200 km s$^{-1}$ with respect to the systemic
velocity of Mrk~231, but this identification is doubtful since the
expected stronger N~I $\lambda$1200 transition is not clearly detected
(although blending with the highly blueshifted Ly$\alpha$ emission
could hide some N~I $\lambda$1200 emission). Another possibility is
Fe~ II $\lambda$1144 blueshifted by $-$2750 km s$^{-1}$. The broad
feature at 1375 \AA\ could plausibly be a blend of Ni~II $\lambda$1317
and C~I $\lambda$1329, but with peculiar shifts of about +500 and
--2200 km s$^{-1}$ with respect to systemic, respectively.  We
consider this identification unlikely because (1) these lines are not
seen in I~Zw~1 and (2) given the ionization potentials and abundances
of Ni and C, many other lines should also be present in the COS
spectrum if those IDs were correct, but they are not detected. Highly
blueshifted ($-$3410 km s$^{-1}$) C~II $\lambda$1334.5 is perhaps a
more likely possibility. The 1296 \AA\ feature could conceivably be
residual N~V $\lambda$1243 (+ highly redshifted Ly$\alpha$) emission
affected by strong N~V BAL, but the N~V + Ly$\alpha$ + continuum
emission over $\sim$1270--1290 \AA\ would have to conspire with the
N~V BAL to produce a nearly featureless pseudo-continuum and explain
the absence of the N~V $\lambda$1238 doublet line, which seems very
unlikely. Highly blueshifted ($-$4000 km s$^{-1}$) Si~II $\lambda$1260
may also contribute to this feature. Similarly, the weaker feature
near 1339 \AA\ may be a blend of O~I $\lambda$1302 and Si~II
$\lambda$1304 blueshifted by $-$4500 km s$^{-1}$. These large
blueshifts would be unusual for quasars, but as we discuss below, they
would be similar to the shift measured in the Ly$\alpha$ emission
feature. A similar blueshift ($-$3000 km~s$^{-1}$) has been reported
in C~III] $\lambda$1909, but this feature is badly blended with Fe~III
lines at shorter wavelengths  (Smith et al. 1995; Gallagher et
al. 2002). The C~IV emission feature in the data of Gallagher et
al. (2002) also seems blueshifted, but the blueshifted C~IV absorption
feature complicates the analysis. We return to this point in \S 4.3.

The Ly$\alpha$ emission feature of Mrk~231 is very unusual (Fig. 2).
It is centered at $\sim$1252 \AA\ or 1201 \AA\ in the rest-frame, {\em
  i.e.}  blueshifted by $\sim$ 3500 km s$^{-1}$ with respect to rest
(assuming no contribution from Si~II $\lambda\lambda$1190, 1193, N~I
$\lambda$1200, and Si~III $\lambda$1207 emission). It has a very
peculiar profile with a narrow peak centered at $\sim$1261 \AA\
($\sim$ --1500 km s$^{-1}$ in the restframe) and wings extending over
at least 1240 -- 1265 \AA\ ($\sim$6000 km s$^{-1}$) and quite possibly
as much as 1225 -- 1280 \AA\ ($\sim$13,000 km s$^{-1}$). The broad
width clearly points to an AGN origin for this feature. The Ly$\alpha$
emission is centered on the NUV nucleus and spatially unresolved with
the rather coarse spatial resolution of COS at 1250 \AA\ for our
settings ($\sim$1$\farcs$3; Ghavamian et al. 2010). The integrated
Ly$\alpha$ flux is $\sim$4.9 $\times$ 10$^{-14}$ ergs s$^{-1}$
cm$^{-2}$ in the line core and 8.5 $\times$ 10$^{-14}$ if one includes
the broad wings (with corresponding rest equivalent widths of 29.2 and
46.6 \AA, respectively; see Table 1). In comparison, the H$\alpha$
flux measured from our 2012 KPNO spectrum is $\sim$140 $\times$
10$^{-14}$ ergs s$^{-1}$ cm$^{-2}$ (the same within the uncertainties
as that measured from our older ESI spectrum).  The
Ly$\alpha$/H$\alpha$ flux ratio of Mrk 231 is therefore only
$\sim$0.5\% of the high density case B value of about 13.  Collisional
suppression of the level 2 population in hydrogen is known to be
important in the high-density BELRs of quasars but reduces the
Ly$\alpha$/H$\alpha$ ratio by at most a factor of $\sim$2 -- 3
relative to the Case B value (e.g., Netzer et al. 1995).  The
integrated Ly$\alpha$/H$\alpha$ ratio in Mrk~231 is thus $\la$2\% the
values typically measured in BLRs. This already small ratio is further
reduced by at least an order of magnitude if we consider only the
H$\alpha$ emission in the core of the line compared to Ly$\alpha$ at
the same velocity (see Figure 4 and discussion in \S 4.2). As we
discuss in the next section, plausible explanations are Ly$\alpha$
destruction by dust grains and the possibility that the observed
H$\alpha$ and Ly$\alpha$ lines form in different regions.

\section{Discussion}

\subsection{The FUV Continuum Emission}

Figure 3 shows the SED of Mrk~231 from $\sim$1000 \AA\ to $\sim$10,000
\AA, where we have combined the new COS spectrum with archival FOS
spectra and our 2001 Keck optical spectrum. Our COS spectrum matches
up precisely with the Nov.\ 1996 G160H spectrum from Gallagher et
al. (2002) without rescaling; there is therefore no evidence for FUV
variability. The older G190/G270H observations from Nov.\ 1992 from
Smith et al. (1995) are systematicaly lower, perhaps by as much as a
factor of 2, than our spectra and those of Gallagher et al.; this
indicates real FUV variations and/or lost flux in the older pre-COSTAR
observations. In Figure 3, we rescaled the Smith et al. G270H spectrum
to join smoothly with the Gallagher et al. data.  At optical
wavelengths, a comparison of the Keck optical spectrum shown in the
figure with our new KPNO spectrum shows no evidence for variations in
the optical continuum and line emission of Mrk~231. The KPNO spectrum
is not shown in Figure 3 because it does not cover as broad a
wavelength range as the Keck data.

In the following we assume, based on the results from spectroscopic
and spectropolarimetric studies (e.g, Thompson et al. 1980; Schmidt \&
Miller 1985 Goodrich \& Miller 1994; Smith et al. 1995, 2004;
Gallagher et al. 2005), that the bulk of the optical continuum
emission in Mrk~231 is produced by the AGN.  The situation at shorter
wavelengths is less clear.  As first pointed out by Smith et
al. (1995), there is a dramatic change in the SED slope around
$\sim$2400 \AA. As shown in Figure 3, the shape of the FUV-NUV
spectrum, especially below $\sim$1400 \AA, is not consistent with any
reddening curve, particularly those having the 2175 \AA\ dust
absorption feature (e.g., Conroy et al. 2010).  The dramatic drop in
flux from $\sim$4000 \AA\ to $\sim$2500 \AA\ requires $A_V \sim$ 7
mag.\ (or E(B--V) = $A_V / 3.1$ $\sim$ 2.3 mag.; see Fig.\ 3), but the
continuum level shortward of $\sim$2400 \AA\ lies well above the
predictions. Smith et al. (1995) attribute the bulk of the $<$2400
\AA\ continuum emission to unreddened light from $\sim$10$^5$ hot,
young OB stars surrounding the active nucleus.  They argue that this
component dilutes the polarized light from the nucleus, resulting in
the rapid decline in the degree of polarization they measure from
$\sim$3200 \AA\ to $\sim$1800 \AA.

Our COS spectrum provides crucial new constraints on this issue. Hot O
and early B stars are expected to produce several stellar photospheric
and wind features within the rest-frame 1100-1400 \AA:
P~V $\lambda\lambda$1118, 1128, Si~IV $\lambda\lambda$1122, 1128,
C~III $\lambda\lambda$1176, 1247, N~V $\lambda\lambda$1238, 1243, O~V
$\lambda$1371, Fe~V $\lambda\lambda$1360-1380, and Si~IV
$\lambda\lambda$1394, 1403.
(e.g., Leitherer et al. 2001; Robert et al. 2003; V\'azquez et
al. 2004).  The lack of any of these features, especially C~III
$\lambda$1176 (EW $\la$ 0.8 \AA), indicates that hot OB stars cannot
contribute more than $\sim$40\% of the FUV continuum emission. This
number is based on the C~III $\lambda$1176 equivalent width of 2 \AA\
measured by Vazquez et al. (2004) in NGC~1705-1.  This estimate for
the stellar contribution is a conservatively high value since we are
assuming here that the weak inflection at $\sim$1225 \AA\ is entirely due
to C~III $\lambda$1176 {\em i.e.} we neglect the possibility of
kinematic substructures in the Ly$\alpha$ emission mimicking this
inflection.  The apparently large width ($\ga$1000 km s$^{-1}$ FWZI)
of this feature makes this identification as a stellar feature very
unlikely.
Indeed our measurements of Si~IV $\lambda\lambda$1394, 1403 suggest
smaller contributions from hot stars.

We conclude that the AGN is the dominant source of the FUV continuum
emission in our data. Both thermal and non-thermal AGN processes may
in principle contribute to the observed FUV continuum emission.  But,
as discussed in Smith et al. (1995), synchrotron radiation from the
AGN can safely be ruled out based on the fact that the polarization
measurements of the strong emission lines are nearly the same as that
of the nearby continuum. We thus favor the scenario where the FUV
light is thermal emission from a geometrically thin, optically thick
accretion disk (e.g., Davis, Woo, \& Blaes 2007 and references
therein). The slowly declining FUV continuum (consistent with
$F_\lambda \propto \lambda^{1.7}$, within the COS wavelength range)
and the near-zero polarization at $\sim$1800 \AA\ (Smith et al. 1995)
implies that this emission is affected only slightly by dust ($A_V
\sim 0.5$ mag.). This puts strong constraints on the dust distribution
as we discuss in \S 4.3.

\subsection{Broad and Blueshifted Ly$\alpha$ Emission}

The large blueshift of the broad Ly$\alpha$ emission in Mrk~231
($\sim$ --3500 km s$^{-1}$) is unusual among AGN.  Most data on broad
emission-line shifts are based on the C~IV $\lambda\lambda$1548, 1550
doublet rather than Ly$\alpha$ since C~IV is less affected by the
intervening IGM (Ly$\alpha$ forest), is free of nearby line emission
(N~V blends with Ly$\alpha$ at low resolution), and is accessible from
the ground starting at lower redshifts than Ly$\alpha$.  The mean
shift of C~IV among radio-quiet quasars like Mrk~231 is --800 km
s$^{-1}$ (--300 for radio-loud systems; Richards et al. 2011). A shift
of --3500 km s$^{-1}$ falls well beyond the distribution of C~IV
blueshifts measured by Richards et al. (their figure 2, which is based
on $\sim$30,000 SDSS quasars). We come to a similar conclusion when
using Ly$\alpha$ directly, although here the statistics are
considerably weaker (Kramer \& Haiman 2009, based on only $\sim$80
quasars): the median blueshift of Ly$\alpha$ in these low-redshift
unobscured quasars is of order --400 km s$^{-1}$ and, according to
their Figure 5, a shift of --3500 km s$^{-1}$ is seen in only two
other (unnamed) objects.  The large Ly$\alpha$ blueshift in Mrk~231 is
even more unusual when we consider only FeLoBAL QSOs. In these
objects, the broad absorption trough removes the blueshifted
Ly$\alpha$ emission, resulting in a Ly$\alpha$ emission profile that
is shifted to the red (e.g., Reichard et al. 2003; Trump et al. 2006;
Gibson et al. 2009; Richards et al. 2011). The only objects with
blueshifted emission lines similar to Ly$\alpha$ in Mrk~231 are the
``PHL 1811 analog'' quasars, where C~IV is blueshifted by at least
$-$6000 km~s$^{-1}$ and tends to have lower-than-average equivalent
width (e.g., Wu et al. 2011, 2012).

To gain insight into the origin of this large Ly$\alpha$ blueshift, it
is instructive to compare the Ly$\alpha$ emission profile of Mrk~231
with the features detected at optical wavelengths.  The results from
this comparison are summarized in Figure 4, where the unscaled
continuum-subtracted emission profiles of Ly$\alpha$ and H$\alpha$ and
absorption profile of Na I~D derived from our recent optical spectrum
are displayed on the same velocity scale (the profiles of the other
Balmer emission lines are not shown here because they are much less
reliable due to lower S/N and strong Fe II emission affecting H$\beta$
and H$\gamma$). The difference between the profiles of H$\alpha$ and
Ly$\alpha$ is stunning. Within the uncertainties of the continuum
placement, the profile of the H$\alpha$ emission is symmetric,
centered on the systemic velocity of Mrk~231 and with broad wings
extending to at least $\pm$7500 km s$^{-1}$ from systemic and a slight
excess emission around $-$5000 km s$^{-1}$ (see also Punsly \& Lipari
2005).

The Ly$\alpha$ emission profile is a better match to the absorption
profile of Na I~D than to the emission profile of H$\alpha$. While the
Ly$\alpha$ emission extends to redder wavelengths than the Na I
absorption, both of these features are highly blueshifted with blue
wings extending to at least --8000 km s$^{-1}$. This comparison
suggests that the small fraction of the Ly$\alpha$ emission which is
able to escape from Mrk~231 does so preferentially along the flow
direction of the FeLoBAL gas (e.g., radially away from the central
engine; see Hall et al. 2004 for a similar explanation of the
peculiar UV Fe~II spectrum of SDSS J091103.49+444630.4). This can
happen naturally if the emission comes from an outflow region with a
significant radial velocity gradient (e.g., Hamann et al. 1993). If
the outflowing BAL systems were instead in front of the source of the
Ly$\alpha$ emission, we would expect the Ly$\alpha$ emission to be
suppressed, rather than preferentially transmitted, at the blueshifted
velocities corresponding to the foreground Ly$\alpha$ BAL (e.g.,
Verhamme, Schaerer, \& Maselli 2006; Schaerer et al. 2011).  This is
seen for instance in $z \sim 3$ Lyman Break Galaxies, where the
Ly$\alpha$ emission is redshifted by approximately twice the outflow
velocity measured from interstellar absorption lines (e.g., Shapley et
al. 2003). Moreover, the BALR would have to cover the BELR without
covering the FUV continuum source -- otherwise one would have to tune
the BAL optical depths to get transmitted fluxes nearly equal to the
continuum at adjacent wavelengths -- but we already know from the
optical spectra (e.g., Rupke et al. 2002) that the BAL gas covers at
least 80-90\% of the optical continuum source (it is a lower limit
because light from an old stellar population may also contribute to
the residual light in the Na I D trough; Smith et al. 1995).  To avoid
these difficulties, we favor a scenario where most, if not all, of the
observed Ly$\alpha$ emission is produced in the outflowing material
itself and is not part of the standard BELR; this scenario is
discussed next.

\subsection{Physical Model}

The data on Mrk~231 are complex with different wavelengths and
features each providing different clues on the geometry of the central
engine. Figures 5 and 6 show two simple pictures that attempt to
explain the main features of these data, particularly the peculiar
Ly$\alpha$ emission, the slowly declining and nearly unpolarized FUV
continuum, and the absence of broad absorption features in the FUV
despite the presence of a strong optical-NUV FeLoBAL outflow. These
simple models are no doubt simplifications of the reality and may not
explain all of the details in the observations. In both pictures, the
outflowing BALR acts as a partial screen for the continuum emission
and as the source of the observed Ly$\alpha$ emission. We discuss
these pictures separately, although the possibility that Mrk~231 is a
hybrid of both cannot be ruled out.

Figure 5 (hereafter referred to as the ``spherical geometry'' or
``spherical scenario'') is an attempt to explain the data using the
simplest geometry possible: the standard BELR is a thick spherical
shell surrounded by a thin dusty shell representing the outflowing
BALR.  The source of the FUV continuum is an accretion disk on scale
of $\sim$few $\times$10$^{15}$ cm ($\la$0.01 pc) with arbitrary
orientation. If this geometry is to explain BAL QSOs in general,
particularly their low frequency of occurrence among QSOs, the BALR
shell must subtend only a small fraction of 4 $\pi$ steradians or be
present for only a fraction of the QSO lifetime (e.g., merger-driven
evolutionary scenario of Sanders et al. 1988; see also Veilleux et
al. 2009; note that the covering factor may be much larger among
dust-reddened quasars like Mrk~231 and those selected via a
FIRST-2MASS cross-correlation, e.g., Urrutia et al. 2009; Glikman et
al. 2012). In Mrk~231, this shell has a small dust-free ``hole'' along
our line of sight to the FUV source that lets 5-10\% of the FUV
continuum through the shell without being attenuated, but a negligible
fraction of the BELR emission goes through this hole since the
projected area of the BELR is much larger than that of the UV
source. In this picture, all of the Ly$\alpha$ emission detected in
Mrk~231 is produced in the dusty BALR shell. We return to this point
below.

Figure 6 (hereafter called the ``disk geometry'' or ``disk scenario'')
is more physically motivated but also more complicated. It is inspired
by Gallagher et al. (2002; their figure 12) and models where the BELR
is comprised of two connected components, the extended accretion disk
atmosphere and a radiation pressure driven wind farther out (e.g.,
Murray et al. 1995; Murray \& Chiang 1995; Proga et al. 2000;
Chelouche \& Netzer 2003; Proga \& Kallman 2004; Proga 2007; Proga,
Ostriker, \& Kurosawa 2008; Richards et al. 2011; Kruczek et al. 2011;
Richards 2012). In these models, the relative emission line
intensities from these two components depend on the physical
conditions in the outflow and therefore, perhaps, the ionizing
spectral energy distribution (SED), as seen by the atmosphere of the
accretion disk.  The BELR and BALR are comprised of clouds / filaments
of various sizes, optical depths, dust content, and ionization
structures (e.g., dusty cold core + hot extended halo); only broad
trends in some key physical properties are shown in Figure 6.  The
boundaries between the BELR and BALR are gradual and probably much
more irregular than shown in this figure.  As illustrated in Figure 6
and described below, the optical-UV spectra seem to favor a line of
sight to Mrk~231 that skirts the upper edge of the dusty wind.

As discussed in \S 4.2, the peculiar blueshifted Ly$\alpha$ emission
in Mrk~231 appears to be direct evidence for line emission associated
with some of the nuclear outflowing material.  The Ly$\alpha$ emission
from a large covering factor outflowing optically thick (in the Lyman
continuum) shell is predicted to be an order of magnitude or more
larger than the flux we observe. Thus, only a few percent of the line
photons need to escape on the blue side of the line to agree with the
observations. This suggests two different ways that are conveyed by
the two scenarios shown in Figures 5 and 6.  The first, illustrated in
Figure 5, is a large covering factor dusty shell containing very low
ionization gas (low ionization parameter) and undergoing large
differential expansion. The dust in this shell absorbs most of the
locally emitted Ly$\alpha$ photons. The only line photons that
reach the observer are the ones escaping from the part of the shell
moving in this direction. The second scenario, illustrated in Figure
6, is a dust-free shell with a small covering factor and/or a small
Lyman continuum optical depth.  In this case the intrinsically
produced Ly$\alpha$ line is very weak and most of it escapes the gas.

A key issue in Figures 5-6 is the location of the dust.  As mentioned
earlier, the spectropolarimetric study of Smith et al. (1995) has
reported near-zero polarization at $\sim$1800 \AA\ which indicates
that the FUV continuum emission is not significantly affected by the
dust that is causing the optical continuum polarization. The dust
distribution must therefore have holes or be patchy, covering only a
fraction of the FUV continuum source. Given the small size of this
source, {\em i.e.}  the inner accretion disk (nominally $\la$ 0.01 pc
across), the dust patches and ``holes'' between the patches must
themselves be very small and, therefore, they are likely to reside
near the UV source rather than in the extended host galaxy (this is
also the conclusions of the spectropolarimetric studies using
different arguments).

At optical wavelengths, we know that a very low ionization BAL outflow
(with substantial Na I, whose ionization potential is only 5.1 eV)
covers at least 80-90\% of the nuclear optical continuum source along
our lines of sight (e.g., Rupke et al. 2002).  A few simple
experiments with the photoionization code {\em CLOUDY} indicate that
dust is not only likely present in this BALR, it seems to be {\em
  required} to shield Na I from the intense UV ($\ga$ 5 eV or
$\la$2400 \AA) radiation field of the AGN that would otherwise
photoionize it to Na II.  So, in both Figures 5 and 6, the dusty BALR
may be the patchy dust absorber causing the optical continuum
polarization (see also Rudy et al. 1985 and Goodrich \& Miller 1994
for independent supportive arguments). If the FUV continuum source is
also not completely covered by the dusty BALR, then the net
observational result could be just what we observe in Mrk~231: a red
partly polarized optical continuum where at least 80-90\% of the flux
is transmitted through the dusty BALR plus a blue nearly unpolarized
FUV continuum spectrum dominated by the small fraction of the FUV
source {\em not} covered substantially by the dusty BALR. This
explanation for the blue FUV spectrum is strongly supported by the
absence of neutral and low-ionization BALs in the FUV, because the
observed FUV flux is not transmitted through this part of the BALR.

Both suggested scenarios are not without difficulties.  In the
spherical geometry of Figure 5, one concern is that a single hole that
transmits only $\sim$5-10\% of the FUV continuum emission may not be
able to explain the considerable amount of unpolarized residual light
in the troughs of the Na I D absorption (10-20\% of the continuum at
$\sim$6000 \AA, depending on aperture size; e.g., Rupke et
al. 2002). The optical continuum source in a thin accretion disk is
predicted to be about an order of magnitude larger than the source of
the FUV continuum (size $\sim$ $\lambda^{4/3}$), so a significantly
smaller fraction of the optical continuum should make it out through
the single hole.  A patchy absorber could explain this result if the
gaps or holes between patches are typically smaller than any continuum
source, but this patchy absorber would also let $\sim$5-10\% of the
BELR emission through the shell without being attenuated, which seems
inconsistent with the small Ly$\alpha$/H$\alpha$ emission ratio.
However, there are a number of issues with these arguments. First, as
mentioned earlier, light from an old stellar population may contribute
significantly to the residual flux in the Na~I BAL troughs.  This
would alleviate the problem by reducing the fraction of the optical
continuum emission from the accretion disk that gets through the
BALR. The fact that the fraction of residual light in the Na I troughs
is lower when using smaller apertures is consistent with this
idea. The slightly positive slope ($\sim$ $\lambda^{1.7}$) of the FUV
continuum, attributed to reddening with $A_V \sim 0.5$ mag., predicts
a larger contribution of (nearly) unpolarized light at optical
wavelengths than in the FUV, which would help explain the optical
data.  Finally, recent microlensing determinations of continuum source
sizes in quasars (e.g., Blackburne et al. 2011) suggest that the
source of continuum emission is larger by nearly an order of magnitude
than the predictions from the standard thin disk models and nearly
independent of wavelengths ($\sim \lambda^{0.2}$ instead of
$\lambda^{4/3}$). The simple single-hole model may therefore be able
to explain the low-polarization FUV continuum, the faint residual
light in the Na~I troughs, and the small Ly$\alpha$/H$\alpha$ emission
ratio all at once.

A second, perhaps more serious, concern with the spherical scenario,
where a single hole transmits nearly all of the observed FUV continuum
emission, is the apparent lack of significant UV continuum variability
in Mrk~231 over a span of at least 15 years (based on the 1996 FOS and
2011 COS data). The hole will change shape and eventually move
across our line of sight to the small FUV continuum source, resulting
in large changes in the amount of FUV emission reaching the
observer. But the timescale for these changes depends on several
factors (e.g., radius of the BALR shell, size of the hole, and density
contrast between the shell and the hole), which can only be explored
with a proper theoretical treatment that is beyond the scope of the
present paper.  A third concern with Figure 5 has to do with the need
for the BALR to redden the NUV-optical continuum by $A_V \sim 7$ mag
and to produce detectable BAL features in not only Na~I and Ca~II H
and K but also He~I$^*$ $\lambda$3889 (which arises from a metastable
excited state populated by recombination from He~II; Leighly,
Dietrich, \& Barber 2011). This suggests the need for neutral and
moderately-ionized columns log~$N_H \ga$ 10$^{22}$ cm$^{-2}$. Again,
it is hard to tell without a proper theoretical treatment if enough
Ly$\alpha$ photons are able to escape from this thick shell.

In the disk geometry of Figure 6, the dusty BALR is in the densest and
most radiatively-shielded environment at the base of the flow.  This
inner BALR extinguishes our view of the BELR and, therefore, helps to
suppress the broad emission lines in the FUV. Dust emerging from the
disk might further suppress emission from the BELR, especially for a
strong resonance line like Ly$\alpha$. The weak blueshifted Ly$\alpha$
emission profile that we do observe could be exempt from this BALR
extinction if it forms a little farther out in the BALR, above the
inner dusty region (see Fig.\ 6). In this picture, the observed
Ly$\alpha$ emission comes from the near-side portion of hot dust-free
BALR, while the Balmer lines arise primarily from the standard BELR
seen through the dusty BALR, in addition to contributions from the
dust-free BALR and possibly also the dusty BALR (hence explaining the
large H$\alpha$/Ly$\alpha$ emission-line ratio and a more symmetric
H$\alpha$ profile). The observed lack of UV continuum variability in
Mrk~231 is a weaker concern in this picture because part of the UV
continuum emission reaches the observer through the dust-free BALR
while the rest comes through the holes (rather than a single hole) of
the patchy Na I BALR. A potentially more serious problem is the
implied orientation of the disk. In the idealized disk geometry of
Figure 6, the absence of redshifted Ly$\alpha$ emission in Mrk~231
(except perhaps for the faint emission feature at $\sim$1296 \AA, see
\S 3, Table 1) implies a rather edge-on view of the disk with $i \ga
45^\circ$ so that the far-side portion of the dust-free BALR is
largely obscured by the dusty BALR on the near-side. On the other
hand, the results from mm-wave / infrared studies on larger few 100 pc
scales indicate that Mrk~231 has a disk viewed rather face on ($i = 10
- 20^\circ$, e.g., Downes \& Solomon 1998; Davies et al. 2004). These
results can be reconciled if this 100-pc scale disk is tilted
significantly with respect to the accretion disk.  The presence of a
warp has in fact been inferred in the inner $\sim$200 pc by Davies et
al. (2004), and results on smaller scales (e.g., Carilli, Wrobel, \&
Ulverstad 1998 and Klockner, Baan, \& Garrett 2003; $i = 56^\circ$)
seem to suggest a more edge-on orientation that provides support for
this picture.  In reality, the boundaries between the dusty and
dust-free BALRs are probably much more irregular than shown in that
figure, so the constraint on the orientation may be relaxed. Moreover,
the ``puffiness'' of the dusty BALR, which sets the covering factor of
the FeLoBAL outflow, is poorly constrained but likely high in
Mrk~231. The derived fraction of LoBALs among quasars depends on the
methods of selection, ranging from only a few percents among UV- or
optically-selected quasars to more than 50\% in the FIRST-2MASS red
quasars, more akin to Mrk~231 (e.g., Urrutia et al. 2009; Glikman et
al. 2012).  Also as mentioned above, the line emission from the BALR
could be mostly in the forward (radial) direction if there are
sufficient velocity gradients to create this avenue for escape. Since
our line of sight is not aligned with the gas flow in the far-side
dust-free BALR, it may be that very little emission from this region
makes it out in our direction. Finally, it is also possible that the
line emission from the far-side BELR and dust-free BALR is attenuated
not only by the foreground dusty BALR but also by other intervening
gas, including the hitchhiking gas which is nearly Compton thick at
least along some sight lines (Braito et al. 2004; Morabito et
al. 2011).

The weak but non-zero C~IV BAL detected by Gallagher et al. (2002) is
expected in both scenarios. In the disk geometry, it happens naturally
along the line of sight that goes through the hot dust-free BALR. In
the spherical geometry of Figure 5, this feature may be explained if
the dust-free hole is filled with ionized gas of lower density (higher
ionization) that has presumably leaked out of the shell.  The lack of
obvious blueshifted Si~IV $\lambda\lambda$1394, 1403 BAL in the
presence of weak C~IV BAL is also not surprising since Si~IV has a
lower ionization potential and lower elemental abundance (in the Sun)
than C~IV, and it is not uncommon to see C~IV BALs accompanied by weak
or absent Si~IV in HiBALs, i.e. the more highly ionized absorbers
(e.g., Trump et al. 2006; Gibson et al. 2009; Allen et al. 2011).  The
lack of N~V $\lambda\lambda$1238, 1243 with C~IV is perhaps harder to
explain in this context since the ionization potential of N$^{+4}$ is
77 eV and the N~V transitions are generally present in HiBALs,
although noticeably weaker than C~IV. As discussed in \S 3 (Table 1),
the emission feature at 1296 \AA\ may conceivably be residual N~V (+
redshifted Ly$\alpha$) emission affected by broad N~V absorption, but
this scenario is a bit contrived. Perhaps a more plausible explanation
is that the C~IV HiBAL detected in the FOS spectra obtained in 1996
has since disappeared without affecting the Na I BAL. This sort of
HiBAL variability would not be unusual (e.g., Gibson et al. 2008; the
0.5 -- 2 keV and 2 -- 10 keV fluxes of Mrk~231 have increased by more
than 50\% from 2001 to 2011, Piconcelli et al. 2012).  New COS
spectroscopy covering Ly$\alpha$ and C~IV $\lambda\lambda$1548, 1550
simultaneously would easily settle the issue.

\section{Summary and Final Remarks}

We have recently obtained a high S/N-ratio FUV spectrum of Mrk~231
covering $\sim$1100 -- 1450 \AA\ as part of our spectroscopic survey
of nearby QSOs with {\em HST}-COS. The spectrum presents faint, broad,
and highly blueshifted Ly$\alpha$ emission. The FUV continuum emission
is slow declining at shorter wavelengths and shows no sign of stellar
photospheric or wind features and no obvious broad absorption line
troughs like those seen in the NUV-optical spectrum of this FeLoBAL
QSO.  These data are best explained if the FUV continuum emission is
produced by the accretion disk of the AGN and the observed Ly$\alpha$
emission is produced in the outflowing BAL cloud systems. The BAL
clouds act as a patchy screen for the FUV continuum emission and as a
polar scattering agent to reproduce the distinct optical polarization
signatures in this object.  Many of these results may be explained
using a simple spherical geometry or a physically motivated but more
complex disk geometry (Figs.\ 5-6).  Both of these scenarios are not
without problems, however.  A proper theoretical treatment beyond the
scope of the present paper will be needed to decide which scenario, if
any, is favored over the other.

One may wonder why other FeLoBAL QSOs do not show the same complex
UV-optical spectrum as Mrk~231. The detection of the Na~I BAL in
Mrk~231 is an important clue since it requires dust. In the disk
scenario, Mrk~231 may be an extreme case of the classic BAL outflow /
dusty torus model of Konigl \& Kartje (1994) with larger column
density, higher mass loss rate, hence more shielding, than usual. This
extra shielding allows dust to survive at smaller radii. This
facilitates the appearance of dust (and Na~I) in the BALR, as well as
the formation of a dusty toroidal wind farther out. This extreme
outflow also leads to more broad line emission from a wind component
of the BELR (e.g., Reichard et al. 2003; Richards et al. 2011). The
amount of covering by this dusty BALR, which is inferred from the
depths of the Ca~II and Na~I troughs (Rupke et al. 2002; Smith et
al. 1995), is consistent with the amount of line-of-sight covering
needed to explain the residual FUV continuum, roughly 80-90\%.

Note finally that in both the spherical and disk geometries of Figures
5 -- 6 the BALR is located within the nuclear $<$1-10 pc scale region
rather than on galactic kpc scales. The location of the BALR is an
important issue since the outflow energetics scale with distance of
the BALR from the central engine and thus have a direct bearing on the
role of quasar feedback in the evolution of galaxies.  Recently,
Faucher-Gigu\`ere et al. (2012) modeled FeLoBAL outflows as ISM clouds
(on $\sim$kpc scales) that are shocked, shredded, and accelerated by a
hot (normal HiBAL) quasar wind. This is inspired by the observational
work of, e.g., Moe et al. (2009), Bautista et al. (2010), and Dunn et
al. (2010) indicating that some of those flows, or some (narrow line)
components of them, do exist at $>$kpc distances.  This model likely
does not apply to Mrk~231 for the following reasons: 1) The Na~I BALs
of Mrk~231 are not spatially resolved on sub-kpc scale (Rupke et
al. 2002; Rupke \& Veilleux 2011, 2012).  2) The Na~I BALs varied, at
least in the highest velocity components. This puts Mrk~231 in line
with several other BAL variability studies and, in particular, with
the work by Hall et al. (2011) on a variable FeLoBAL, that favors
pc-scale origins for the BALs. 3) The detached high-velocity Na~I
troughs in Mrk~231, which imply a lack of nuclear absorbers with
velocities between $\sim$--3500 and 0 km s$^{-1}$, are difficult to
explain in the models of Faucher-Gigu\`ere et al. (2012), where the
cold gas is accelerated by the quasar blast wave starting at the
near-zero velocity of the ambient ISM.  4) Several spectropolarimetric
studies of Mrk~231 have suggested that the FeLoBAL clouds are causing
the optical continuum polarization, which is due to dust scattering in
the polar illumination cones of the circumnuclear region, roughly
along the pc-scale jet axis of Mrk~231. This result would be hard to
explain if the FeLoBAL clouds were distributed on galactic scale,
where the influence of the central engine is much more isotropic
(Rupke \& Veilleux 2011, 2012).  While these arguments favor a nuclear
origin for the FeLoBAL system in Mrk~231, none of them rule out the
possibility that this type of shocked and accelerated ISM model may
apply to the spatially resolved galactic-scale outflow in this object
(Rupke \& Veilleux 2011, 2012; Faucher-Gigu\`ere \& Quaetert 2012).

\acknowledgements Support for this work was provided to S.V., M. T.,
and T. M. T. by NASA through contracts {\em HST} GO-1256901A and
GO-1256901B.  S.V. also acknowledges support from a Senior NPP Award
held at the NASA Goddard Space Flight Center, where most of this paper
was written, and from the Humboldt Foundation to provide funds for a
long-term visit at MPE in 2012.  F.H. acknowledges support from the
National Science Foundation through grant AST-0908910.  We thank the
referee, Patrick Hall, for a thorough report and thoughful suggestions
which improved the paper. S.V., F.H., and D.L. also thank A. Laor and
H. Netzer for organizing an excellent AGN meeting in Haifa, where many
of the ideas presented here germinated. This work has made use of
NASA's Astrophysics Data System Abstract Service and the NASA/IPAC
Extragalactic Database (NED), which is operated by the Jet Propulsion
Laboratory, California Institute of Technology, under contract with
the National Aeronautics and Space Administration.

\clearpage

\begin{deluxetable}{lclcccc}
\tablecolumns{7}
\tabletypesize{\scriptsize}
\tablecaption{Uncertain Features in the COS Spectrum}
\tablewidth{0pt}
\tablehead{
\colhead{$\lambda_{\rm obs}$ (\AA)} & \colhead{$\lambda_{\rm rest}$ (\AA)} & 
\colhead{$EW$ (\AA)} & \colhead{$F_{em}$ (10$^{-15}$ erg s$^{-1}$ cm$^{-2}$)} & \colhead{Possible IDs} & \colhead{Velocity (km s$^{-1}$)} & \colhead{Confidence} \\
\colhead{(1)} & \colhead{(2)} & \colhead{(3)} & \colhead{(4)} & \colhead{(5)} & \colhead{(6)} & \colhead{(7)} 
}
\startdata
1181.3 (0.8) & 1133.5 & $-$3.3 (0.3) & 3.6 (0.8) & N~I $\lambda$1134.4 & $-$240 & 3\\
     & &     & & Fe~II $\lambda$1144.0 & $-$2750 & 3\\
1225.4 (0.1) & 1175.8  & $+$0.8 (0.2) & ... & C~III $\lambda$1175.6 & $+$50& 3\\
1252.0 (0.4) & 1201.3 & $-$29.2\tablenotemark{a} / $-$46.6\tablenotemark{a} & 49\tablenotemark{a} / 85\tablenotemark{a} & Ly$\alpha$ $\lambda$1215.7 & $-$3550 & 0\\
1252.1 (0.05) & 1201.4 & $+$0.13 (0.02) & ... & Ly$\alpha$ $\lambda$1215.7  & $-$3530 & 1\\
1296.0 (0.3) & 1243.5 & $-$3.6 (0.8) & 6.4 (1.4) & Ly$\alpha$ $\lambda$1215.7& $+$6900 & 3\\
     & &     & & N~V $\lambda$1242.8 & $+$170 & 3\\
     & &     & & Si~II $\lambda$1260.4 UV4 & $-$4020 & 3\\
1324.0 (0.2) & 1270.4 & $-$2.6 (0.5) & 2.7 (0.5) & ??? & ??? & 3\\
1339.5 (0.5) & 1285.2 & $-$4.1 (1.2) & 4.7 (0.9) & O I $\lambda$1302.2 + Si~II $\lambda$1304.4 & $-$4200 & 3\\
1355.2 (0.1) & 1300.3 & $+$2.7 (0.6) & ... & O I $\lambda$1302.2 + Si~II $\lambda$1304.4 & $-$740 & 3\\
1375.0 (0.7) & 1319.3 & $-$8.2 (2.0) & 13.0 (1.7) & Ni II $\lambda$1317.2 & $+$480 & 3\\
     & &     & & C~I $\lambda$1328.8 & $-$2140 & 3\\
     & &     & & C~II $\lambda$1334.5 UV1 & $-$3410 & 3\\
1391.8 (0.1) & 1335.4 & $+$1.8 (0.2) & ... & C~II $\lambda$1334.5 UV1 & $+$200 & 2\\
             &        &      & ... & C~II$^*$ $\lambda$1335.7 UV1 & $-$70 & 2\\
1452.0 (0.4) & 1393.2 & $+$0.7 (0.1) & ... & Si~IV $\lambda$1393.8 & $-$130 & 2\\
1462.0 (1.0) & 1402.8 & $+$0.2 (0.2) & ... & Si~IV $\lambda$1402.8 & $-$0 & 2\\
\enddata
\tablecomments{ Col.\ (1): Wavelength of the feature (centroid) in the
  observer's frame in \AA. Col.\ (2): Wavelength of the feature
  (centroid) in the rest frame ($z$ = 0.0422) in \AA.  Col.\ (3):
  Equivalent width in \AA\ measured in the observer's frame (negative
  value corresponds to emission).  Col.\ (4): flux of the emission
  feature in 10$^{-15}$ erg s$^{-1}$ cm$^{-2}$. Col.\ (5): Possible
  line identification based on, e.g., Morton 2003; Laor et al. 1997;
  Zheng et al. 1997, vanden Berk et al. 2001; Hall et al. 2002.  Col.\
  (6): Velocity of the feature (centroid) corresponding to the
  tentative line identification.  Col.\ (7): Level of confidence of
  the line identification: 0 = certain, 1 = confident, 2 = uncertain, 3 =
  unlikely. See \S 3 for more detail.}

\tablenotetext{a}{The first number corresponds to the core of the
  Ly$\alpha$ emission line, while the second number also includes the
  contribution from the broad wings.}

\label{tab:tab1}
\end{deluxetable}

\clearpage

\begin{figure*}
\epsscale{0.7}
\centering
\includegraphics[width=0.85\textwidth,angle=90]{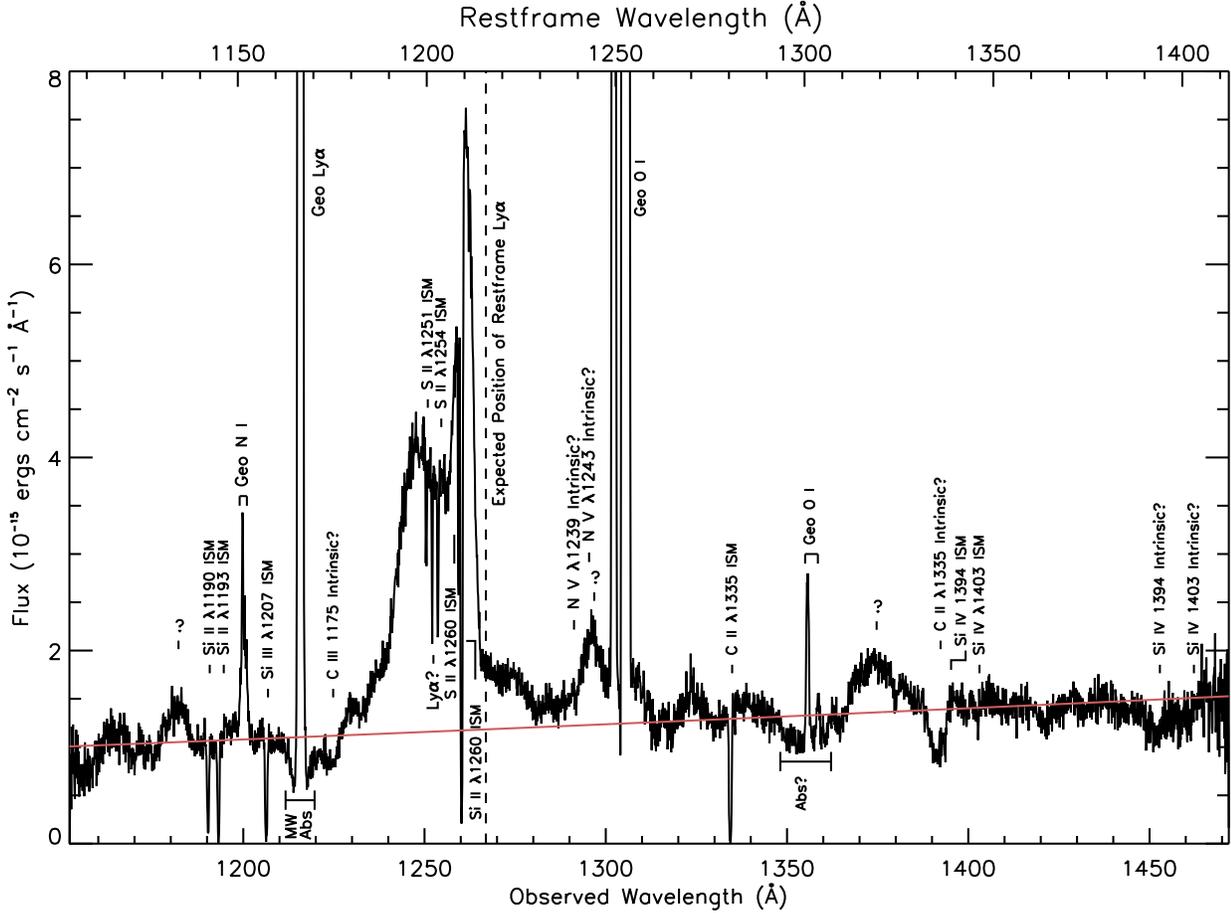}
\caption{{\em HST}-COS FUV spectrum of Mrk~231, binned by 10 spectral
  pixels to help show the fainter features. It has not been corrected
  for the small foreground Galactic extinction ($A_V \sim 0.03$ mag.).
  This spectrum is dominated by broad, highly blueshifted Ly$\alpha$
  emission. The FUV continuum emission is nearly featureless and only
  slowly declining at shorter wavelengths, consistent with $F_\lambda
  \propto \lambda^{1.7}$ (shown as a red line in the figure). It is
  thus dominated by the AGN and only slightly affected by dust
  reddening ($A_V \sim 0.5$ mag.) from the neutral and dusty broad
  absorption line region. }
\label{fig:cos_spec}
\end{figure*}

\begin{figure*}
\epsscale{0.7}
\centering
\includegraphics[width=0.8\textwidth,angle=90]{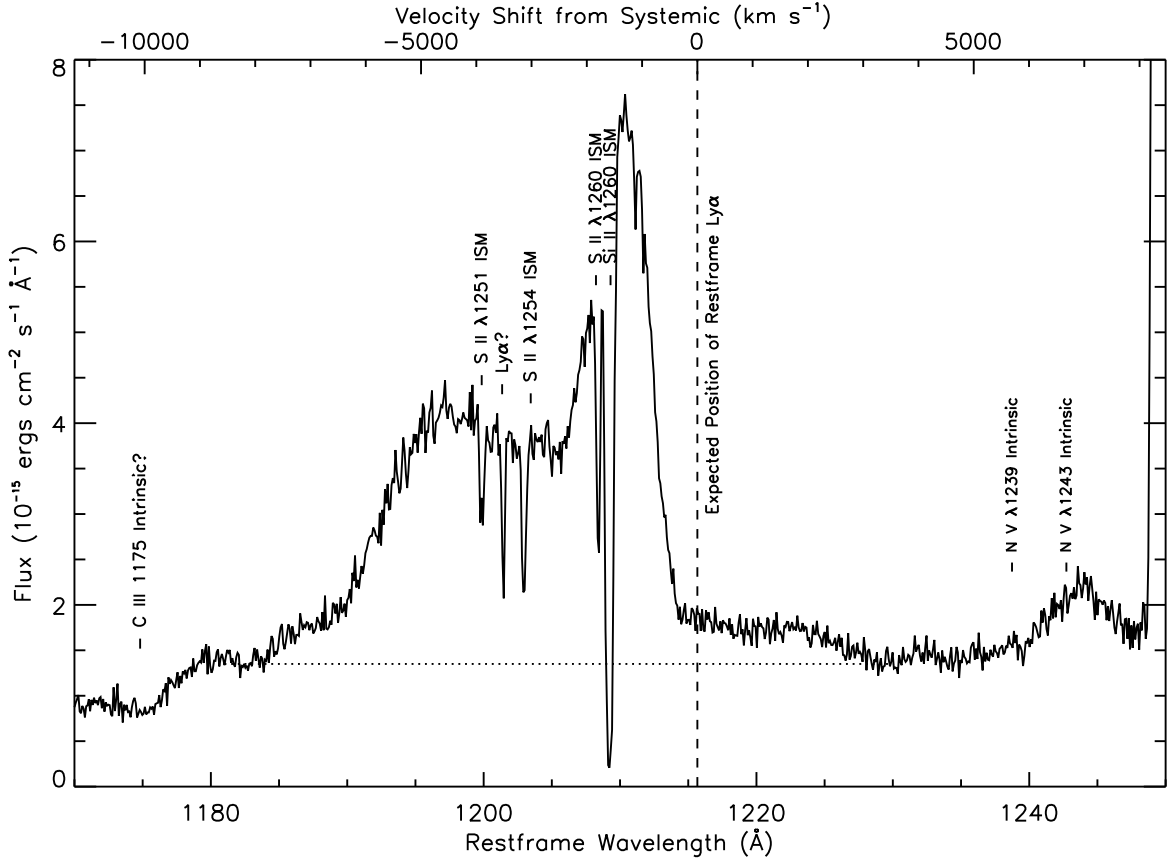}
\caption{Ly$\alpha$ emission profile in Mrk~231. The vertical dash
  line indicates the expected position of rest-frame Ly$\alpha$, while
  the dotted line indicates a conservatively high estimate of the
  continuum level used to produce Figure 4.  Note the large blueshift
  of the Ly$\alpha$ emission with respect to systemic (the centroid of
  the profile is at $\sim$ --3500 km s$^{-1}$) and the width of the
  line ($\ga$10,000 km s$^{-1}$ at the base of the profile) which
  indicates an AGN origin. There is no indication that Si~II
  $\lambda\lambda$1190, 1193, N~I $\lambda$1200, and Si~III
  $\lambda$1207 are significantly affecting the profile of
  Ly$\alpha$. The feature at restframe wavelength 1243.5 \AA\ is
  unidentified, but it could plausibly be highly redshifted Ly$\alpha$
  emission at $\sim$+6900 km s$^{-1}$ (see \S 3 and Table 1 for more
  detail).}
\label{fig:cos_spec}
\end{figure*}

\begin{figure*}
\epsscale{0.5}
\centering
\includegraphics[width=0.75\textwidth,angle=90]{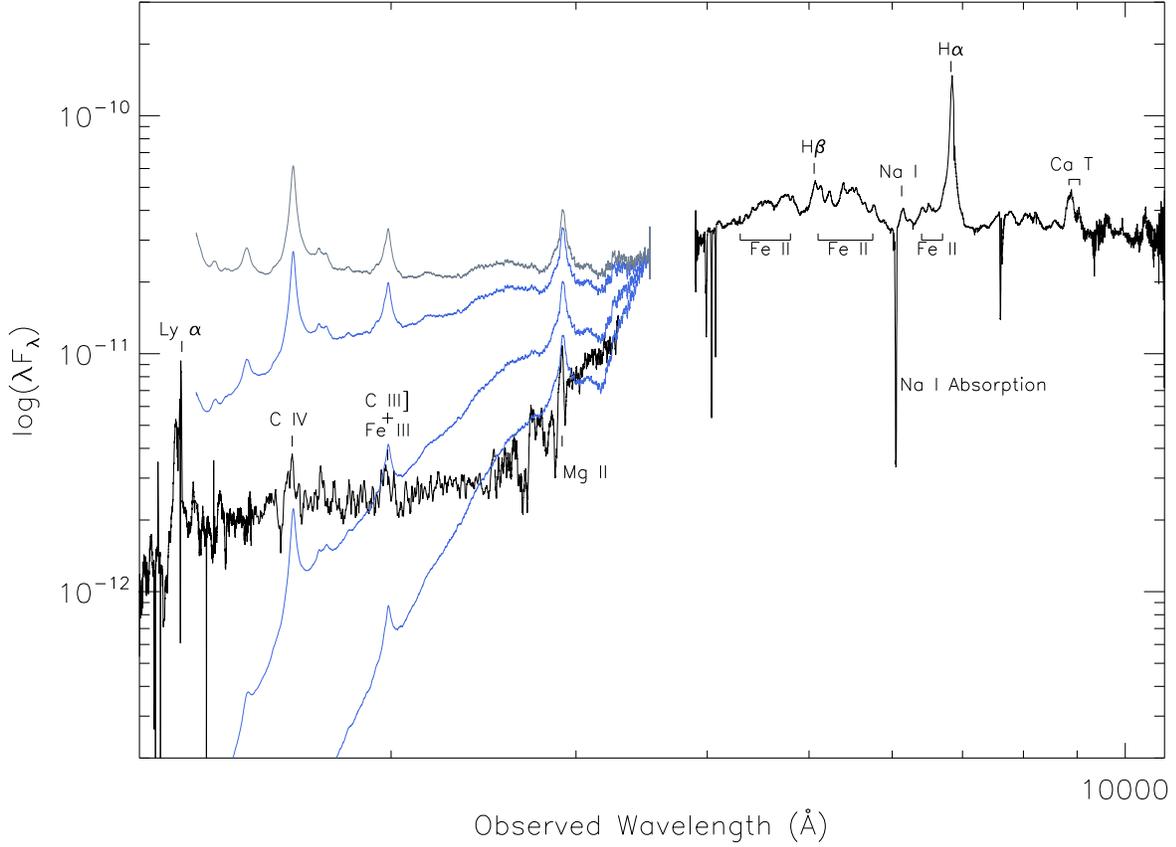}
\caption{ The UV -- optical spectral energy distribution (SED) of
  Mrk~231 in log $\lambda f_\lambda$ {\em versus} log
  $\lambda$(observed) units.  This SED combines the new COS data with
  archival FOS spectra and the 2001 Keck/ESI optical spectrum
  published in Rupke et al. (2002). The old pre-COSTAR FOS NUV
  spectrum at 2500 -- 3300 \AA\ was scaled up to match the data at
  shorter wavelengths (see \S 4.1 for more detail). A comparison with
  our new KPNO spectrum shows no evidence for variations in the
  optical continuum and line emission of Mrk~231. The KPNO spectrum is
  not shown here because it does not cover as broad a wavelength range
  as the Keck data.  The blue curves indicate the {\em HST}/FOS
  composite quasar spectrum used in the {\em HST} COS ETC (shown in
  grey) reddened by $A_V$ = 1.0, 4.0, and 7.0 mag.\ using $R_V = A_V /
  E(B - V)$ = 3.1 and an extinction curve without the 2175 \AA\ dust
  absorption feature (Conroy et al. 2010), normalized to match the
  observed spectrum of Mrk~231 at $\sim$3600 \AA.  Note the dramatic
  drop in flux in Mrk~231 from $\sim$4000 \AA\ to $\sim$2500 \AA,
  requiring dust reddening $A_V \sim 7$ mag., followed by a slowly
  declining continuum below that wavelength, only slightly affected by
  dust ($A_V \sim 0.5$ mag.). This is consistent with the near-zero
  polarization at $\sim$1800 \AA\ found by Smith et al. (1995). }
\label{fig:sed}
\end{figure*}

\begin{figure*}
\epsscale{0.7}
\centering
\includegraphics[width=0.75\textwidth,angle=90]{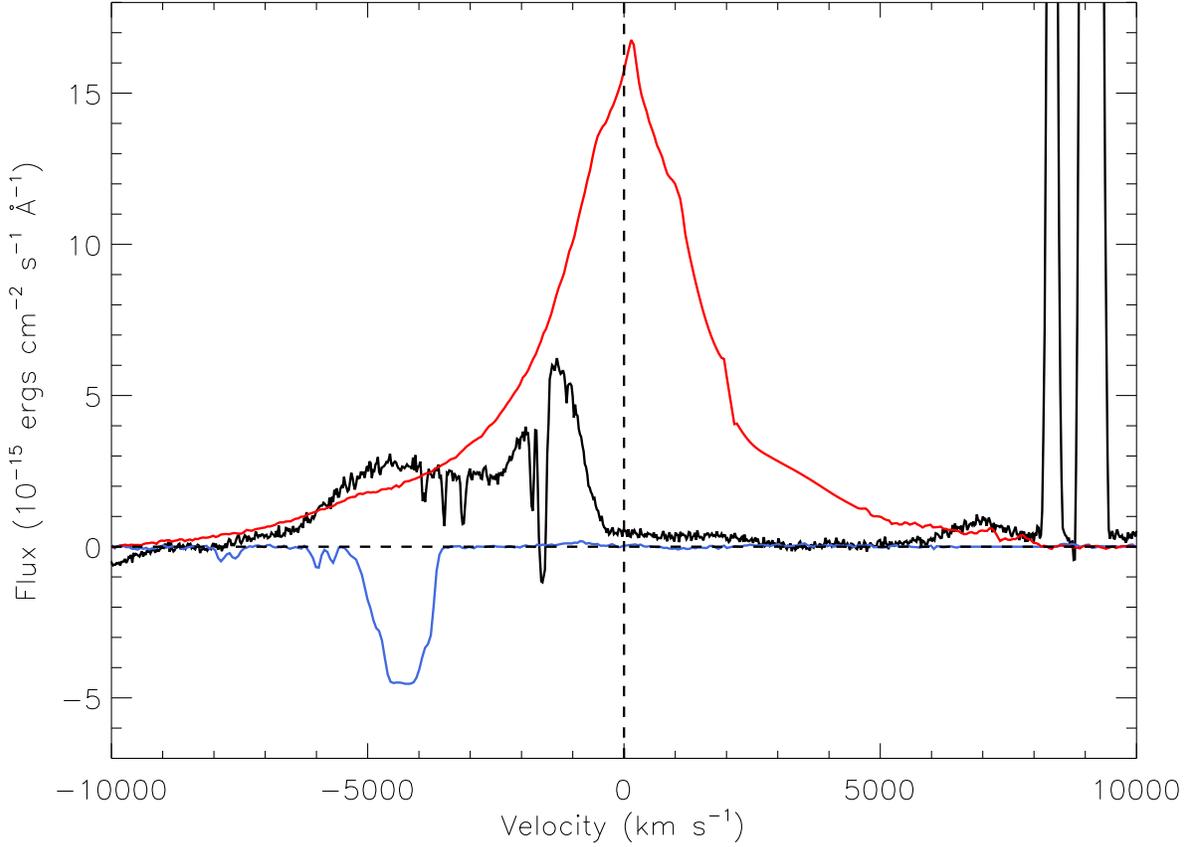}
\caption{ Comparison of the continuum-subtracted H$\alpha$ emission
  (red), Ly$\alpha$ emission (black), and Na I~D absorption
  (blue). The conservatively high continuum level shown in Figure 2
  was used to derive the Ly$\alpha$ emission profile shown here. The
  optical features are derived from our most recent KPNO data to
  reduce profile variability. All three features are on the same
  velocity and absolute flux scales. The H$\alpha$ profile was
  interpolated at $\sim$1000 -- 2000 km s$^{-1}$ to remove the
  telluric O$_2$ $\lambda$6850 band. Ly$\alpha$ shares a stronger
  resemblance with Na~I~D absorption than with H$\alpha$ emission. The
  emission feature at $\sim$ +6900 km s$^{-1}$ could conceivably be
  faint redshifted Ly$\alpha$ emission (see \S 3 and Table 1 for more
  detail).}
\label{fig:comparison}
\end{figure*}

\begin{figure*}
\epsscale{1.0}
\centering
\includegraphics[width=0.9\textwidth,angle=0]{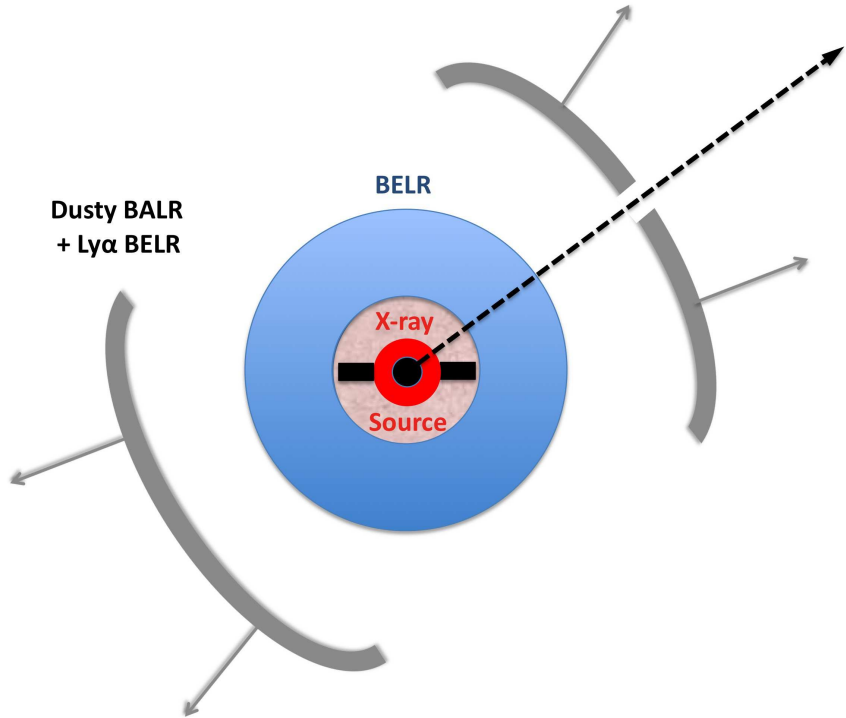}
\caption{ A simple geometric spherical model of the central pc-scale
  regions of Mrk~231 (not drawn to scale).  The BELR (in blue) is modeled 
  as a thick shell
  while the BALR (in grey) is a thin, differentially expanding dusty shell with
  a small hole along the line of sight to the central FUV-emitting
  accretion disk ($\la$0.01 pc, shown in black). This small hole 
  lets only a small fraction ($\sim$5-10\%) of the FUV continuum
  source through the shell. BELR clouds along the line of sight and
  hot gas (shown in pink) within the inner radius of the BELR, fully
  ionized by the X-ray source (shown in red), contribute to the X-ray
  measured Compton-thick absorbing column density.  In this picture,
  the observed FUV continuum emission is the $\sim$5-10\% that
  goes through the hole unabsorbed. All of the BELR emission, produced
  on larger scale, is either completely absorbed by the shell (FUV resonance
  lines; e.g., Ly$\alpha$) or attenuated (NUV-optical lines).  The
  weak C~IV BAL measured by others suggests that the hole in the shell
  is actually filled with dust-free ionized material of lower density (higher 
  ionization).
  Most 
  of the Ly$\alpha$ produced in the shell itself
  is absorbed on the way out partly by direct absorption and partly due 
  to photon scattering on the grains. 
  The small fraction of the Ly$\alpha$ photons that makes it out of
  the shell comes from the part of the shell facing the observer,
  resulting in the observed large blueshift (see \S 4.3 for more
  detail).  }
\label{fig:comparison}
\end{figure*}

\begin{figure*}
\epsscale{1.0}
\centering
\includegraphics[width=0.75\textwidth,angle=0]{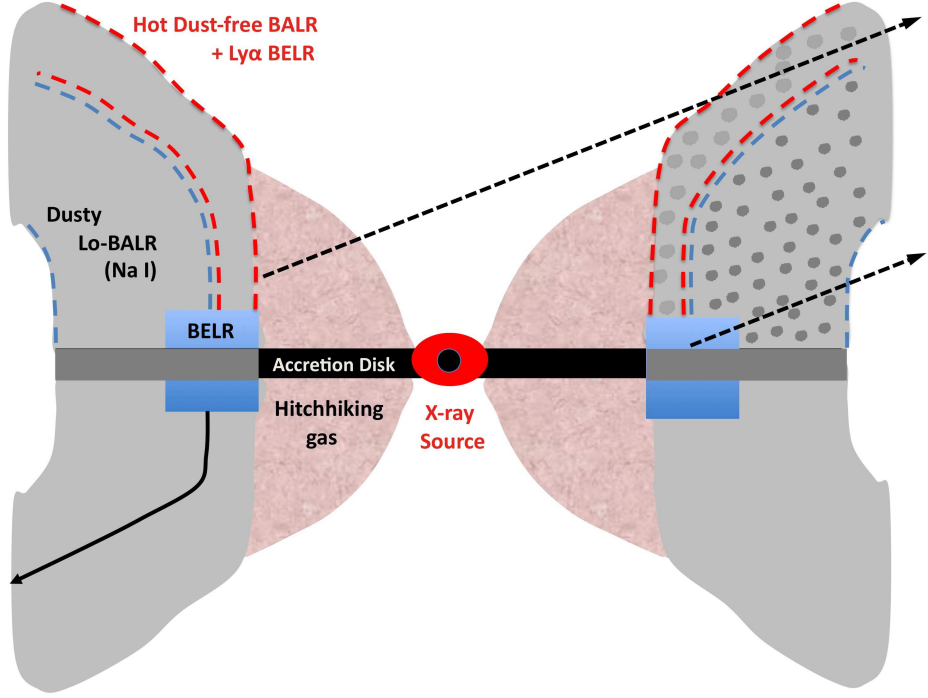}
\caption{ A physically motivated geometric disk model of the central
  pc-scale regions of Mrk~231 (not drawn to scale).  A strong clumpy
  outflow (shown in grey, with a typical flow line shown in solid
  black) is present, lifted off the accretion disk (in dark grey) and
  radiatively accelerated by the UV radiation field from the
  UV-emitting accretion disk (in black). Gas too close to the
  central X-ray source (shown in pink and dubbed the ``hitchhiking
  gas'' by Murray et al. 1995) is too ionized and thus too transparent
  to UV radiation to take part in this wind; it does however
  contribute significantly to the X-ray absorbing column density.  The
  portion of the outflow which is closest to the UV and X-ray source
  is hot and free of dust; it is responsible for the weak C~IV BAL
  feature and some of the broad line emission, particularly the
  blueshifted Ly$\alpha$. The densest and most radiatively-shielded
  environment at the base of the flow is neutral and dusty; it is
  responsible for the strong Na I BAL seen in the optical. The
  boundaries between the various regions labeled in this figure are
  gradual and probably much more irregular than shown here.  Two lines
  of sight are shown for illustration. The top one goes through the
  hot dust-free ($A_V \sim 0$ mag.) BALR, where all of the observed
  Ly$\alpha$ emission, most of the C~IV emission, and some Balmer line
  emission are produced. This line of sight is not aligned with the
  gas flow in the far-side dust-free BALR, so very little Ly$\alpha$
  emission from this region makes it out in our direction.  The FUV
  continuum detected along this line of sight is seen through the hot
  dust-free BAL, producing the faint C~IV BAL. The lower line of sight
  goes through the patchy dusty BALR ($A_V \sim 7$ mag.) where most of
  the optical continuum passes through the Na~I BALR clouds (hence it
  is reddened and partly polarized). The small fraction of the optical
  and UV continuum that does not pass through the Na~I BALR clouds is
  unpolarized; this accounts for only a small fraction of observed
  optical continuum, but perhaps a significant fraction of the
  observed FUV flux (see \S 4.3 for more detail).  }
\label{fig:comparison}
\end{figure*}

\clearpage

\end{document}